\documentclass[aps,prl,twocolumn,twoside,showpacs]{revtex4}
\usepackage{amsfonts}
\begin{document}

\title{
Right-handed Neutrino Fields are Real Spinors}
\author{ Xin-Bing Huang}
\email[]{huangxb@pku.edu.cn} \affiliation{Department of Physics,
Peking University, Beijing 100871, the People's Republic of China}
\date{June 28, 2005}

\begin{abstract}
The ansatz that the right-handed neutrino fields are real spinors
is proposed in this letter. We naturally explain why the
right-handed neutrinos don't feel the electroweak interactions and
why there is neutrino mixing. It is found that the Majorana
representation of Dirac equation is uniquely permitted in our
scenario. With this ansatz, we predict that: 1. there are at least
four species of neutrinos; 2. the mass matrix of neutrinos must be
traceless; 3. there is CP violation in the lepton sector. The
difference between our scenario and the Zee model is discussed
also.

\end{abstract}

\pacs{14.60.St, 14.60.Pq, 12.15.Ff, 11.30.Rd}
\maketitle

Till now there are enormous experiments on weak interactions.
About half a century ago, Lee and Yang~\cite{ly56} proposed the
violation of parity conservation in weak interactions. A short
time after that, the experiments~\cite{wah57} revealed that only a
single helicity appears in weak interactions: electrons and
neutrinos are always created left-handed, positrons and
antineutrinos are always created right-handed. Recently there are
abundant data~\cite{ahm02,soi03,apo99,LSND} from atmospheric,
solar, underground laboratory, and long baseline neutrino
experiments on the neutrino mass and mixing. All of these
experimental data, except for the results observed by LSND
collaboration~\cite{LSND}, can be explained by oscillations
between three active neutrinos~\cite{neuosc}. A light {\em
sterile} neutrino has been assumed to interpret the LSND data.
Neutrino oscillations provide direct evidence of nonzero neutrino
masses and mixing between different mass eigen states of
neutrinos. A massive neutrino must exist in both left-handed and
right-handed states~\cite{xin04}. Furthermore, left-handed
neutrinos traversing a strong gravitational field could be
converted into right-handed ones~\cite{cp91}. Now that
right-handed neutrinos must exist in universe, why don't they feel
both the weak interactions and the electromagnetic interactions?

Now it is believed that massive neutrinos are either Dirac or
Majorana~\cite{maj37} particles. But neither Dirac nor Majorana
fields can embody the essential difference between the left-handed
and right-handed neutrinos. In quantum field theory, the local
gauge invariance, the spin and the charge possessed by a quantum
field is deeply related with the algebra describing this field.
For instance, the $\pi^{0}$, being a neutral spin-0 boson, is
characterized by a real scalar, whereas $\pi^{+}$ and $\pi^{-}$,
being charged spin-0 bosons, have to be represented by complex
scalar. In the stand electroweak model~\cite{gsw60}, we have to
represent the charged intermediate vector bosons as follows
\begin{equation}
\label{y0034} W^{\pm}_{\mu}=\frac{1}{\sqrt{2}}(B^{1}_{\mu}\mp
iB^{2}_{\mu})~,
\end{equation}
where $B^{1}_{\mu}$ and $B^{2}_{\mu}$ are real SU(2) gauge fields
introduced by Yang and Mills~\cite{yan54}, whereas the $Z^{0}$ is
described by a real vector. Hence a simple and natural ansatz to
get out of above mentioned dilemma is that the right-handed
neutrino fields are real spinors.

The left-handed and right-handed fields for a spinor $\psi$ are
defined by
\begin{eqnarray} \label{chiral}
\psi_{L}=\frac{1}{2}(1-\gamma_5)\psi~,~~~~\psi_{R}=\frac{1}{2}(1+\gamma_5)\psi~.
\end{eqnarray}
In this case, our ansatz can be represented by
\begin{eqnarray} \label{realcom}
\psi_{L}=\frac{1}{2}(\psi-\psi^{*})~,~~~~\psi_{R}=\frac{1}{2}(\psi+\psi^{*})~.
\end{eqnarray}
Obviously, comparing Eq.(\ref{realcom}) with Eq.(\ref{chiral})
directly yields
\begin{eqnarray} \label{rela1}
\gamma_5\psi=\psi^{*}~.
\end{eqnarray}
When $\psi$ is the mass eigen state of a fermion, then it is easy
to prove that Eq.(\ref{rela1}) can be satisfied if and only if
that: 1. the Majorana representation of Dirac equation is adopted;
2. this fermion is massless and uncharged.

In the case of four generations of neutrinos, each of four
neutrino mass eigen states shall be represented by $\nu_{i}$;
$i=1,2,3,4$. Because of neutrino mixing, we have the linear
superposition
\begin{equation}
\label{mix}\nu_{\alpha} =\sum_{i=1,2,3,4}U_{\alpha i}\nu_{i}~,
\end{equation}
where $U_{\alpha i}$ is a unitary matrix, $U^{\dag}U=1$, and
$\alpha=e,\mu,\tau,s$ represents the weak flavor eigen states
(corresponding to electron, muon, tauon and {\em sterile}
neutrinos respectively), namely
\begin{eqnarray}\label{simplify}
\nu=\left( \begin{array}{c } \nu_{1} \\ \nu_{2} \\ \nu_{3}\\
\nu_{4}
\end{array} \right)~,
~~~~~~~ \tilde{\nu}=\left( \begin{array}{c } \nu_{e} \\ \nu_{\mu} \\
\nu_{\tau} \\
\nu_{s}
\end{array} \right)~.
\end{eqnarray}
Then Eq.(\ref{mix}) can be compactly rewritten as $\tilde{\nu}
=U\nu$.

When there is neutrino mixing, our ansatz is thus expressed in
mathematical language as
\begin{eqnarray} \label{reladd}
\begin{array}{l}\displaystyle
\nu_{\alpha
R}=\frac{1}{2}(1+\gamma_{5})\nu_{\alpha}=\frac{1}{2}(\nu_{\alpha}+\nu_{\alpha}^{*})~,
\\[0.3cm]
\displaystyle
 \nu_{\alpha
L}=\frac{1}{2}(1-\gamma_{5})\nu_{\alpha}=\frac{1}{2}(\nu_{\alpha}-\nu_{\alpha}^{*})~,
\end{array}
\end{eqnarray}
or equivalently
\begin{eqnarray} \label{rela2}
\gamma_5\nu_{\alpha}=\nu_{\alpha}^{*}~.
\end{eqnarray}
Treating the ansatz as the first principle in this letter, we
investigate the possible Lagrangian which make sure that the
constraint of Eq.(\ref{rela2}) is satisfied.

Consider the kinetic energy term of neutrinos in electroweak
Lagrangian as follows($\hbar=c=1$)
\begin{eqnarray}\label{dlang1}
 {\cal L}_{k}(\nu)&=&\frac{i}{2}\left( {\bar
{\nu}}\gamma^{\mu}\partial_{\mu}\nu+ {\bar
{\nu}^{*}}\gamma^{\mu}\partial_{\mu}\nu^{*}
\right)
\\
\nonumber &=&\frac{i}{2}\sum_{\alpha=e,\mu,\tau,s}\left( {\bar
{\nu}_\alpha}\gamma^{\mu}\partial_{\mu}\nu_\alpha+{\bar
{\nu}_\alpha^{*}}\gamma^{\mu}\partial_{\mu}\nu_\alpha^{*}
\right)~.
\end{eqnarray}
It is obvious that the kinetic energy term keeps invariant under
the transformation of Eq.(\ref{rela2}). In the condition of
Eq.(\ref{rela2}), the above Lagrangian is represented in terms of
left-handed and right-handed neutrino fields as follows
\begin{eqnarray}\label{dlang2}
{\cal L}_{k}(\nu)=i
 {\bar {\tilde{\nu}}_{L}}\gamma^{\mu}\partial_{\mu}\tilde{\nu}_{L}
 +i{\tilde{\nu}_{R}^{T}}\gamma_{0}\gamma^{\mu}\partial_{\mu}\tilde{\nu}_{
 R}
~,
\end{eqnarray}
where ${\bar {\tilde{\nu}}_{L}}\equiv
{\tilde{\nu}}_{L}^{\dag}\gamma_{0}$, and ${\tilde{\nu}_{R}^{T}}$
denotes $(\nu_{e R}^{T},\nu_{\mu R}^{T},\nu_{\tau R}^{T},\nu_{s
R}^{T})$. Obviously the second term in the right hand side of
above Lagrangian indicates that the right-handed neutrino fields
can not keep both U$_{Y}(1)$ and SU$_{R}(2)$ local gauge
invariance by introducing the SU$_{R}(2)\times$U$_{Y}(1)$ gauge
fields. Hence our ansatz nicely explains why the right-handed
neutrinos don't feel the electroweak interactions.

To investigate the antineutrinos, we introduce the charge
conjugate fields
\begin{equation}
\label{antip} \tilde{\nu}_{c}=C\tilde{\nu}^{*}~,
\end{equation}
where $C$ is the charge conjugation matrix, satisfying
\begin{equation}
\label{ccm}
C^2=1~,~~C^{\dag}=C~,~~C\gamma_{\mu}^{*}C=-\gamma_{\mu}~.
\end{equation}
After above definitions on the charge conjugation, our ansatz on
neutrinos, namely Eq.(\ref{rela2}), implies that
\begin{equation}
\label{ans2} \gamma_{5}\tilde{\nu}_{c}= -\tilde{\nu}^{*}_{c}~,
\end{equation}
because $C\gamma_{5}^{*}C=-\gamma_{5}$. This relation demonstrates
that the left-handed antineutrinos defined by
\begin{equation}
\label{left-anti} \tilde{\nu}_{c
L}=\frac{1}{2}(1-\gamma_{5})\tilde{\nu}_{c}=\frac{1}{2}(\tilde{\nu}_{c}+\tilde{\nu}^{*}_{c})
\end{equation}
don't feel the electroweak interactions either.

We have demonstrated that our ansatz doesn't break the
SU$_{L}(2)\times$U$_{Y}(1)$ symmetry in the sector of left-handed
neutrinos. The condition Eq.(\ref{rela2}) keeps massless
electroweak Lagrangian invariant, but our ansatz will give
stringent constraints on neutrino mass matrix and its mixing
matrix.

Assume that the $\nu_{1}$, $\nu_{2}$, $\nu_{3}$, and $\nu_{4}$
correspond to four mass eigen states of masses $m_1$, $m_2$,
$m_3$, and $m_4$ respectively, then the addition of a mass term to
total Lagrangian which is possibly invariant under the
transformation of Eq.(\ref{rela2}) should be
\begin{eqnarray}\nonumber
 {\cal L}_{m}&=&-\frac{1}{2} \left(
{\nu}^{\dag}\gamma_{0}M\nu+{\nu}^{*\dag}\gamma_{0}M\nu^{*} \right)
\\
\label{masslag} &=& -\frac{1}{2}\left(
\tilde{\nu}^{\dag}\gamma_{0}U M
U^{\dag}\tilde{\nu}+\tilde{\nu}^{*\dag}\gamma_{0}U^{*} M
U^{*\dag}\tilde{\nu}^{*} \right)
\end{eqnarray}
because $U^{-1}=U^{\dag}$. Where $M$ is the diagonal mass matrix,
satisfying
\begin{eqnarray}\label{massmatrix}
M=\left( \begin{array}{cccc} m_{1}&0&0&0 \\ 0 & m_{2}& 0&0 \\
0&0&m_{3}&0 \\0&0&0&m_4
\end{array} \right)~.
\end{eqnarray}
Making use of Eq.(\ref{rela2}), the relation
$\gamma_{0}\gamma_{5}=-\gamma_{5}\gamma_{0}$, and the identity
$(\gamma_{5})^2=1$, we acquire
\begin{eqnarray}\nonumber
 {\cal L}_{m}&=&\frac{1}{2} \left(\tilde{\nu}^{*\dag}\gamma_{0}U M
U^{\dag}\tilde{\nu}^{*}+\tilde{\nu}^{\dag}\gamma_{0}U^{*} M
U^{*\dag}\tilde{\nu}\right)
\\
\nonumber &=&\frac{1}{2} \left({\nu}^{*\dag}\gamma_{0}U^{*\dag}U M
U^{\dag}U^{*}{\nu}^{*}+ \right.
\\
\label{masslag2} &&~~~~\left.  {\nu}^{\dag}\gamma_{0}U^{\dag}U^{*}
M U^{*\dag}U \nu \right)~.
\end{eqnarray}
Comparing Eq.(\ref{masslag2}) with Eq.(\ref{masslag}), one can
find out that the Lagrangian $ {\cal L}_{m}$ keeps invariant if
and only if the following identity is satisfied
\begin{equation}
\label{masscon} U^{*\dag}U M U^{\dag}U^{*}=-M~.
\end{equation}
Define a new mass matrix
\begin{equation}
\label{newmass} M_{\nu}=U M U^{\dag}~,
\end{equation}
the identity (\ref{masscon}) obviously means that
\begin{equation}
\label{newmass2} M_{\nu}=-M_{\nu}^{*}~.
\end{equation}
Since $M_{\nu}=M_{\nu}^{\dag}$, then $M_{\nu}^{T}=-M_{\nu}$,
therefore, the new mass matrix is imaginary and antisymmetric.
Furthermore, $Tr(M_{\nu})=0$ makes sure that
\begin{equation}
\label{newmass3} Tr(M)=0~.
\end{equation}
Recently the traceless neutrino mass matrix has been
systematically investigated in Refs.\cite{hz03} in the case of
three generations of neutrinos.

We would like to deduce the identity Eq.(\ref{masscon}) from the
viewpoint of Dirac equation again. The Dirac equation for free
neutrinos is taken to be
\begin{equation}
\label{dirac} (i\gamma^{\mu}\partial_{\mu}-M)\nu=0~,
\end{equation}
in terms of $\tilde{\nu}$, the above equation becomes
\begin{equation}
\label{dirac2}
(i\gamma^{\mu}\partial_{\mu}-M)U^{\dag}\tilde{\nu}=0~.
\end{equation}
Multiplying $\gamma_{5}$ in both sides of Eq.(\ref{dirac2}), and
making use of Eq.(\ref{rela2}) yields
\begin{equation}
\label{dirac3}
(-i\gamma^{\mu}\partial_{\mu}-UMU^{\dag})\tilde{\nu}^{*}=0~,
\end{equation}
Taking the complex conjugation for both sides of
Eq.(\ref{dirac2}), we obtain
\begin{equation}
\label{dirac4}
(-i\gamma^{*\mu}\partial_{\mu}-M)U^{*\dag}\tilde{\nu}^{*}=0~,
\end{equation}
There is a representation called the Majorana representation in
which the $\gamma$s are all imaginary, so that
$\gamma^{*\mu}=-\gamma^{\mu}$. Under this representation,
Eq.(\ref{dirac4}) is simplified to
\begin{equation}
\label{dirac5} (i\gamma^{\mu}\partial_{\mu}-U^{*}M
U^{*\dag})\tilde{\nu}^{*}=0~,
\end{equation}
Comparing the above equation with Eq.(\ref{dirac3}) directly
yields the identity (\ref{newmass2}). Here we have also proved
that the Majorana representation of Dirac equation must be
adopted. Our ansatz automatically breaks the freedom of selecting
the representation.

We explicitly write the mass matrix $M_{\nu}$ out
\begin{eqnarray}\label{massmatrix2}
M_{\nu}=i\left( \begin{array}{llll} 0 &  m_{e\mu} & m_{e\tau} & m_{e s}
\\ - m_{e\mu} & 0 & m_{\mu\tau} & m_{\mu s}\\
-m_{e\tau} & -m_{\mu\tau} & 0 & m_{\tau s}\\-m_{e s} & -m_{\mu s}
& -m_{\tau s} & 0
\end{array} \right)~.
\end{eqnarray}
The definition (\ref{newmass}) shows that
\begin{equation}
\label{deter} {\rm det}\mid M \mid={\rm det}\mid M_{\nu}\mid~,
\end{equation}
namely
\begin{eqnarray}\label{massid}
\begin{array}{l}
m_{1}m_{2}m_{3}m_{4}=-2m_{e\mu}m_{\mu\tau}m_{\tau s}m_{e
s}+m_{e\tau}^{2}m_{\mu s}^{2}
\\[0.2cm] ~~~~~~~~~~~~~~~~~~~
-m_{e s}^{2}m_{\mu\tau}^{2}-m_{e\mu}^{2}m_{\tau s}^{2}~.
\end{array}
\end{eqnarray}
The equation (\ref{deter}) obviously indicates that the species of
neutrinos must be no less than four to make sure the real neutrino
masses.

In the Majorana representation, $\gamma_{0}$ is an imaginary
matrix, then we can set
\begin{equation}
\label{gammadef} \gamma_{0}=i\Gamma~,~~~~M_{\nu}=i N~,
\end{equation}
where $\Gamma$ and $N$ are real matrices. Since
$\gamma_{0}^{\dag}=\gamma_{0}$, then
\begin{equation}
\label{antisym} \Gamma^{T}=-\Gamma~,~~~~N^{T}=-N~.
\end{equation}
Expressing the neutrino fields in real and imaginary parts
respectively
\begin{equation}
\label{twocom} {\nu}_{\alpha}=\Psi_{1\alpha}+i\Psi_{2\alpha}~,
\end{equation}
here $\Psi_{1\alpha}$ and $\Psi_{2\alpha}$ being real functions,
then Eq.(\ref{reladd}) is equivalently rewritten as
\begin{equation}
\label{twocomlr} {\nu}_{\alpha
R}=\Psi_{1\alpha}~,~~~~{\nu}_{\alpha L}=i\Psi_{2\alpha}~.
\end{equation}
After these definitions, we rewrite the Lagrangian (\ref{masslag})
as follows
\begin{eqnarray}\nonumber
 {\cal L}_{m}
&=& - \tilde{\nu}^{\dag}\gamma_{0} M_{\nu} \tilde{\nu}=
\sum_{\alpha ,\beta} {\nu}^{\dag}_{\alpha }\Gamma N_{\alpha\beta}
{\nu}_{\beta}
\\
\nonumber &=&\sum_{\alpha ,\beta} \left(\Psi_{1\alpha}^{T}\Gamma
N_{\alpha\beta}\Psi_{1\beta}+\Psi_{2\alpha}^{T}\Gamma
N_{\alpha\beta}\Psi_{2\beta}\right.
\\
\nonumber && \left. +i\Psi_{1\alpha}^{T}\Gamma
N_{\alpha\beta}\Psi_{2\beta}-i\Psi_{2\alpha}^{T}\Gamma
N_{\alpha\beta}\Psi_{1\beta}\right)
\\
\label{masslag4} &=& \sum_{\alpha ,\beta}  \left(
{\nu}^{T}_{\alpha R}\Gamma N_{\alpha\beta} {\nu}_{\beta R}  +
{\nu}^{\dag}_{\alpha L}\Gamma N_{\alpha\beta} {\nu}_{\beta L}
\right)~.
\end{eqnarray}
It is astonishing that there are no left-handed and right-handed
neutrino couplings in above mass term. We thus have systematically
demonstrated that the left-handed neutrinos can independently keep
local gauge invariance.

In the literature, the parametrization of neutrino mixing matrix
has been studied extensively~\cite{lm05}. In our scenario, the
neutrino mixing matrix is a $4\times 4$ unitary matrix, which will
be more complicated. The equation (\ref{masscon}) indicates that
our neutrino mixing matrix can not be real, therefore there is CP
violation in the lepton sector.

In the Zee model~\cite{zee1}, adding a charged +1 Higgs field
which transforms as a singlet under SU(2) to the standard
electroweak model leads to neutrino Majorana masses. The Zee mass
matrix $M_{\nu}$, generated by radiative correction at one loop
level~\cite{zee2}, is symmetric and has null diagonal entries.
Thus the neutrino mixing matrix $V$ is determined by
$M=V^{T}M_{\nu}V$. The confrontation of Zee model with
experimental data~\cite{zee3} and possible extended Zee
models~\cite{zee4} have been extensively investigated. There are
some similarity between the Zee model and our scenario since our
mass matrix (\ref{massmatrix2}) has also null diagonal entries.
But the form of our mass matrix is decided by the algebraic
property of the neutrino fields. Obviously the neutrino described
by our ansatz is neither Dirac nor Majorana particles.
Furthermore, it is easy to incorporate our ansatz into the stand
electroweak model without introducing an extra Higgs field.

In conclusion, after proposing an algebraic symmetry (\ref{rela2})
of neutrino fields, we have naturally explained why the
right-handed neutrinos don't feel the electroweak interactions and
why there exists the neutrino mixing. We have shown that the
Majorana representation of Dirac equation must be adopted. Our
ansatz provide stringent constraints on neutrino mass matrix and
its mixing matrix. The real mass condition needs a light {\em
sterile} neutrino outside three active neutrinos. The complex
mixing matrix has predicted the leptonic CP violation.

\end{document}